# STAGES OF STEADY DIFFUSION GROWTH OF A GAS BUBBLE IN STRONGLY SUPERSATURATED GAS-LIQUID SOLUTION

**Kuchma A. E., Gor G. Yu., Kuni F. M.**

*Institute of Physics, Saint-Petersburg State University*

*198504, Ulyanovskaya str., 1, Petrodvorets, St. Petersburg, Russia*

Gas bubble growth as a result of diffusion flux of dissolved gas molecules from the surrounding supersaturated solution to the bubble surface is studied. The condition of the flux steadiness is revealed. A limitation from below on the bubble radius is considered. Its fulfillment guarantees the smallness of fluctuation influence on bubble growth and irreversibility of this process. Under the conditions of steadiness of diffusion flux three stages of bubble growth are marked out. With account for Laplace forces in the bubble intervals of bubble size change and time intervals of these stages are found. The trend of the third stage towards the self-similar regime of the bubble growth, when Laplace forces in the bubble are completely neglected, is described analytically.

INTRODUCTION

The aim of the present paper is to elaborate a theoretical description of a gas bubble growth during a phase transition in a strongly supersaturated gas-liquid solution. Formation and growth of gas bubbles in a solution is a highly widespread phenomenon both in nature, and in technological processes. Gas bubble growth plays an important role in the production of microcellular materials based on polymers, metals or glass. Besides, growth of water vapor bubbles dissolved in magmatic melt is a process which leads to volcanic eruptions.

The theory of the phenomenon under consideration is formally similar to the theory of cavitation [1, 2] or to the theory of boiling processes in superheated liquid [3, 4]. However, there is a considerable difference between these theories caused by the difference in the physical nature of initial metastable states of the system. Metastability of liquid phase in cavitation or boiling is related, correspondingly, with expanding or with overheating of liquid. During decomposition of liquid solution supersaturated with gas the solvent usually has the same temperature and pressure as the solution, corresponding to the stability area of the liquid phase; and solution metastability is connected with the fact that its concentration exceeds the concentration of saturated solution.

Let us first make an important clarification. After the instant creation of supersaturation in the solution of gas in liquid and after a certain delay a gas bubble nucleates fluctuationally, and then grows irreversibly. The subject under consideration in the current paper is such irreversibly growing bubble. Right after nucleation the bubble has such a small size that Laplace forces have significant impact on the pressure. Although these forces decrease with bubble growth, nevertheless, as it is determined below, they influence the bubble radius time dependence for a relatively long time.

The growth of a gas bubble in supersaturated gas-liquid solution has been studied earlier in papers [5−16]. In the present paper the following problems that have not been discussed before, are considered. The growth regularities analysis is conducted for high values of initial supersaturation necessary for the fluctuational nucleation of the bubble.

The condition guaranteeing the steadiness of the flux of dissolved gas molecules to the bubble surface is revealed. Earlier this condition has only been obtained for the particular case of self-similar regime of bubble growth [16]. The limitation from below on the bubble radius, the fulfillment of which guarantees the smallness of the influence of fluctuations on the bubble growth, and its irreversibility are considered. Basing on this limitation, we introduce the initial condition to the equation describing bubble growth. Within the assumption of the diffusion flux steadiness we reveal three stages of the bubble growth. With account for Laplace forces in the bubble we obtain intervals of the bubble size change and corresponding time intervals for these stages.

The conditions of steadiness of the flux of dissolved gas molecules to the bubble surface during each of the stages are revealed; and it is shown that, as a rule, there is gradual transition in time from steady regime of bubble growth to the essentially non-steady one. The trend towards the third stage to the self-similar regime of bubble growth studied in [16] is described analytically. In [16] there were no limitations on gas solubility, but the influence of Laplace forces was completely neglected. We compare the results for the time dependence of bubble radius obtained using our theory (accounting for Laplace forces) with the predictions of self-similar theory to evaluate the precision of the latter. This comparison is conducted within the limitations on gas solubility guaranteeing the steady growth of the bubble.

Let us note that the influence of Laplace forces on diffusion growth of a bubble under the conditions of steadiness of the diffusion flux has been considered in paper [5]. The same problem in case of non-steady conditions has been discussed in paper [6], where, however, only a numerical solution of the diffusion equation has been given. In papers [5, 6] there high values of supersaturation, when the fluctuational nucleation of a bubble is only possible have not been considered. The question of the choice of the initial condition for the bubble growth equation has not been discussed at all.

# 1. THE BALANCE EQUATION ON THE NUMBER OF GAS MOLECULES IN THE BUBBLE

The state of the solution is stipulated by temperature $T$, pressure $\Pi$ and initial gas concentration $n_0$ (number density of the dissolved gas molecules). Value $n_\infty$ will stand for the concentration of the saturated solution, which at given temperature $T$ and pressure $\Pi$ is characterized by chemical and mechanical equilibrium with the pure gas above the plane contact surface. The solution is presupposed to be diluted. Dissociation and chemical transformations of the dissolved molecules are neglected. Let us define the solution supersaturation $\zeta$ as

$$\zeta \equiv \frac{n_0 - n_\infty}{n_\infty}. \qquad (1.1)$$

After a certain time lag from the instantaneous creation of supersaturation a bubble nucleates fluctuationally in the solution and starts growing irreversibly. The subject of our research is the evolution of a supercritical bubble from its nucleation and until the role of Laplace forces remains essential. We assign the moment of the bubble nucleation as a time $t$ reference point. The bubble radius is denoted as $R$.

The bubble perturbs the solution. Therefore, the initial concentration $n_0$ has the meaning of the concentration at an infinite distance from the bubble. Assuming the mechanical equilibrium between the bubble and the solution, we can write the following expression for the gas pressure in the bubble $P_R$

$$P_R = \Pi + \frac{2\sigma}{R}, \qquad (1.2)$$

where $\sigma$ is the surface tension of the solvent (in the case of diluted solution).

Let us denote the equilibrium concentration of the solution at the bubble surface as $n_R$. Obviously, $n_R$ is the concentration of saturated solution which is in chemical and mechanical equilibrium with the pure gas above the plane contact surface at temperature $T$ and pressure $P_R$. Using Henry's law [17], we have

$$\frac{n_R}{n_\infty} = \frac{P_R}{\Pi}, \tag{1.3}$$

where it is presupposed that the properties of the solvent weakly depend on its pressure at a given temperature. Substituting (1.2) in (1.3), we obtain

$$\frac{n_R}{n_\infty} = 1 + \frac{2\sigma}{\Pi R}. \tag{1.4}$$

Introducing the radius $R_c$ of a critical bubble according to

$$R_c \equiv \frac{2\sigma}{\Pi \zeta} \tag{1.5}$$

(see e. g. paper [7]), we can rewrite (1.4) as

$$\frac{n_R}{n_\infty} = 1 + \zeta \frac{R_c}{R}. \tag{1.6}$$

From (1.6) accounting for (1.1) we have the following equation for the difference between the solution concentration at an infinite distance from the bubble and concentration at the bubble surface $n_0 - n_R$, acting as a driving force of bubble growth:

$$n_0 - n_R = (n_0 - n_\infty)\left(1 - \frac{R_c}{R}\right). \tag{1.7}$$

Since the bubble is supercritical, then $R > R_c$.

Assuming the gas in the bubble to be ideal and accounting for (1.2), we have the following expression for the number of molecules $N$ in the bubble

$$N = \frac{4\pi R^3}{3kT}\left(\Pi + \frac{2\sigma}{R}\right), \tag{1.8}$$

where $k$ is the Boltzmann constant. The equation on the balance of gas molecules number in the bubble requires that

$$\frac{dN}{dt} = -4\pi R^2 j_R, \tag{1.9}$$

where the right hand side is the diffusion flux of dissolved gas molecules towards the bubble surface.

Assuming the dissolved gas concentration profile around the bubble to be steady,

we have the following expression for the diffusion flux density $j_R$:

$$j_R = -\frac{D}{R}(n_0 - n_R), \qquad (1.10)$$

where $D$ is the diffusion coefficient of gas molecules in the liquid solvent (in case of diluted solution). Let us notice that in (1.10) the value of dissolved gas concentration at the bubble surface is presumed to be equal to the equilibrium value. As it is demonstrated in [8, 10], such assumption is justified with the observance of the strong inequality $R \gg 4sD/\alpha_c v_T$, where $s$ is the gas solubility that is defined as a dimensionless value via

$$s \equiv \frac{kTn_\infty}{\Pi}, \qquad (1.11)$$

$\alpha_c$ ($\alpha_c \leq 1$) is the condensation coefficient of a gas molecule at the virtual transition of gas molecules from the bubble and into the solution, and $v_T$ is a mean heat velocity of a gas molecule inside the bubble. At typical values $10^{-2} < s < 10^{-1}$, $D \sim 2\times 10^{-5}\,cm^2 s^{-1}$, $v_T \sim 3\times 10^4\,cm\,s^{-1}$, $R_c \sim 10^{-7}\,cm$ this inequality is justified even at $R = R_c$, unless $\alpha_c$ is not anomalously small. The condition of validity of the steady approximation in (1.10) will be clarified in the end of this section.

Rewriting the balance equation (1.9) using (1.7), (1.8) and (1.10), and accounting for (1.1), we obtain the bubble growth equation

$$\dot{R} = Ds\zeta\left(1 - \frac{R_c}{R}\right)\frac{1}{R}\left(\frac{1}{1 + R_\sigma/R}\right). \qquad (1.12)$$

Here $\dot{R} \equiv dR/dt$; and $R_\sigma$ is the characteristic bubble radius defined as

$$R_\sigma \equiv \frac{4}{3}\frac{\sigma}{\Pi}. \qquad (1.13)$$

In accordance with (1.12) (and $R > R_c$), we have $\dot{R} > 0$, which corresponds to the monotonous growth of the bubble radius with time.

From (1.13) and (1.5) it follows that

$$\frac{R_\sigma}{R_c} = \frac{2\zeta}{3}. \qquad (1.14)$$

Fluctuational nucleation of a gas bubble is possible only in the case of high values of initial supersaturation of the solution, when the following strong inequality is fulfilled

$$\zeta \gg 10. \tag{1.15}$$

Then, (1.14) demonstrates that $R_\sigma \gg R_c$. At the same time from (1.13) one can see that in the interval $R \ll R_\sigma$ (and $R > R_c$) it is that $2\sigma/R \gg \Pi$ (the role of Laplace forces is high), but in the interval $R \gg R_\sigma$ it is that $2\sigma/R \ll \Pi$ (the role of Laplace forces is low).

Let us introduce characteristic time $t_R \equiv R/\dot{R}$, during which the bubble radius increases significantly. From physical considerations it is clear that the condition of steadiness of dissolved gas concentration in the vicinity of the bubble is equivalent to the requirement of the smallness of bubble radius $R$ in comparison with diffusion length $(Dt_R)^{1/2}$. Strong inequality $R \ll (Dt_R)^{1/2}$ or equivalent strong inequality

$$\left(\frac{R\dot{R}}{D}\right)^{1/2} \ll 1 \tag{1.16}$$

is the condition of the validity of steady approximation (1.10), i. e. the condition of steady bubble growth. Using (1.12), we present the condition (1.16) as inequality

$$\left(s\zeta \frac{R-R_c}{R+R_\sigma}\right)^{1/2} \ll 1, \tag{1.17}$$

which is a strong limitation on the solubility $s$ for the case of high values of $\zeta$ that are under our consideration. This limitation becomes stricter with the growth of $R$. Condition (1.17) was formulated previously only for the particular case of self-similar regime of bubble growth [16], when this condition is reduced to $(s\zeta)^{1/2} \ll 1$.

## 2. THREE STAGES OF BUBBLE GROWTH

Let us now proceed to the examination of bubble growth dynamics in the considered supercritical interval of its sizes. The appropriate equation is intentionally written in a rather awkward form (1.12) for the convenience of its analysis. Each of the three co-factors dependent on $R$ emphasized in the right hand side of (1.12) describes its physically different contribution to the dynamics of the supercritical bubble growth process. Co-factor $1 - R_c/R$, increasing with $R$, corresponds to the fast (the scale of change of $R$ is $R_c$) increase of the driving force of the process (the value $n_0 - n_R$) with the growth of $R$. Co-factor $1/R$, decreasing with the growth of $R$, describes, as it is seen from (1.10), the contribution related to the decrease of the gradient of solution concentration near the bubble surface, which decreases the bubble growth rate with the growth of $R$. Finally, co-factor $1/(1 + R_\sigma/R)$, which increases with the growth of $R$, accounts for the counteraction of Laplace forces to the bubble growth, i. e. the fact that the bubble growth is facilitated by the reduction of Laplace forces with the growth of $R$, while other factors are equal (the scale of its change is $R_\sigma$). Notwithstanding the mentioned reduction of counteraction, the resulting contribution of the last two factors always leads to the deceleration of growth with the increase of bubble size. It becomes evident when the bubble growth equation (1.12) is rewritten in the following form:

$$\dot{R} = Ds\zeta \left(1 - \frac{R_c}{R}\right) \frac{1}{R + R_\sigma}. \tag{2.1}$$

Let us consider a non-trivial question of the introduction of initial condition to the bubble growth equation (2.1). In equation (2.1), as in equation (1.12), the influence of fluctuations on the bubble growth is neglected. Meanwhile, the nucleation of a bubble, later growing irreversibly, can in fact occur only as a result of fluctuational overcoming of the activation barrier; the barrier is overcome not by a single bubble, but by an ensemble of bubbles. Due to this reason it is necessary to find the condition of validity of the assumption on the smallness of fluctuation influence on a bubble growth presupposed in equation (2.1).

Let us use the arguments from [18]. As it was shown in [7-9] these arguments are applicable to the problem under consideration at the condition of mechanical equilibrium between the bubble and the solution.

Although the influence of fluctuations is being attenuated with the strengthening of inequality $R > R_c$, it nevertheless is still essential [18] in the area $0 < R - R_c < (kT/4\pi\sigma)^{1/2}$, where the upper bound is an order of magnitude estimate. Indeed, fluctuations can bring bubbles back from this area to the pre-critical area $R < R_c$. In order to neglect the influence of fluctuations on bubble growth it is necessary that the radius of that bubble to satisfies the strong inequality

$$R - R_c \gg (kT/4\pi\sigma)^{1/2}. \qquad (2.2)$$

Irreversible growth of such bubbles is described by equation (2.1), which replaces a more complicated kinetic equation of Fokker-Planck type, accounting for fluctuations.

The estimation of quantity $(kT/4\pi\sigma)^{1/2}$ for water at standard conditions gives the value $\sim 10^{-8} cm$. For $R_c$ under the condition of strong supersaturation ($\zeta \sim 10^3$) the estimation is $R_c \sim 10^{-7} cm$. Evidently, in this case inequality (2.2) is observed well in the area $R \geq 2R_c$, where bubble growth can be *a fortiori* considered regular. That is why the initial condition to equation (2.1) can be chosen in the form of

$$R(t)|_{t=0} = 2R_c. \qquad (2.3)$$

It is clear that the use of equation (2.1) is possible in the area of sizes $R$, a bit less than $2R_c$. Then, on the one hand, strong inequality (2.2) will be satisfied with less precision; and, therefore, the neglect of fluctuations will be less justified. On the other hand, as we will show in section 3 of the current paper, the range of times available for examination using equation (2.1), will then expand insignificantly.

Let us note that during the initial period of bubble growth corresponding to the initial condition (2.3), when $0 \leq R - 2R_c \ll R_c$, the condition of steady growth of a bubble (1.17) will be (as we will see from inequality (2.17)) satisfied; moreover, this will be valid regardless of the values of solution supersaturation, even when the gas solubility is not too small. Such values of solubility are typical for the case when there is no dissociation and chemical transformations of dissolved molecules.

We will call bubbles which radius is $R \geq 2R_c$ far supercritical ones. Evolution in time of a far supercritical bubble takes place rather regularly (without fluctuations). The limitation from below on the bubble radius, at which the influence of fluctuations on the bubble growth is negligible, was not considered in [5, 6].

Before, using the initial condition (2.3), we find the explicit solution of equation (2.1), describing the size growth of a far supercritical bubble as a dependence on time from the moment of its nucleation. Let us consider the change of the character of bubble growth process with the increase of its size. This will let us mark out the representative stages of growth and to determine corresponding characteristic bubble sizes. Duration of the consecutive stages and the character of bubble radius time dependence on each stage will be considered in the next section. Let us notice that the stages of our interest do not have anything in common with the stages of evolution of the whole ensemble of bubbles during the decomposition of liquid solution supersaturated with gas.

From (1.14) and (1.15) it can be seen that $R_\sigma \gg R_c$. Using this inequality, equation (2.1) and boundary condition (2.3) we have that the growth rate of the bubble radius in its dependence from this radius has to reach the maximum value achieved at certain bubble radius $R = R_m$ from the interval $2R_c < R_m < R_\sigma$. Thus it is natural to consider the growth in the following interval of sizes

$$2R_c \leq R \leq R_m \tag{2.4}$$

as a first stage of supercritical bubble evolution, where the determining factor is the increase of the driving force of the growth. At this stage bubble growth goes with the increasing in time rate, reaching its maximum at $R = R_m$.

In order to obtain $R_m$ we will consider the rate of bubble growth as a function of its radius and differentiate both parts of equation (2.1) with respect to $R$:

$$\frac{d\dot{R}}{dR} = Ds\zeta \frac{R_c(R_c + R_\sigma) - (R - R_c)^2}{R^2(R + R_\sigma)^2}. \tag{2.5}$$

The quantity $R_m$ is defined by the extremal condition

$$\left.\frac{d\dot{R}}{dR}\right|_{R=R_m} = 0, \tag{2.6}$$

which with account for (2.5) leads to the following result

$$R_m = R_c + (R_c^2 + R_c R_\sigma)^{1/2}. \tag{2.7}$$

Accounting for the strong inequality $R_\sigma \gg R_c$, we can simplify the obtained expression for $R_m$:

$$R_m \approx (R_c R_\sigma)^{1/2}(1 + (R_c / R_\sigma)^{1/2}) \approx (R_c R_\sigma)^{1/2}. \tag{2.8}$$

In accordance with (2.4) and (2.8), on the first stage of growth we have the following expression for the bubble radius

$$2R_c \leq R \leq (R_c R_\sigma)^{1/2}. \tag{2.9}$$

As it follows from equation (2.1) and initial condition (2.3), the bubble growth rate at the zero time (at $R = 2R_c$) is

$$\dot{R}\bigg|_{R=2R_c} = \frac{Ds\zeta}{2R_\sigma} \frac{1}{1 + 2R_c / R_\sigma} \approx \frac{Ds\zeta}{2R_\sigma}, \tag{2.10}$$

and the maximum value of growth rate reached at the end of the first stage, can be found from (2.1) using equation (2.8) for $R_m$:

$$\dot{R}\bigg|_{R=R_m} \approx \frac{Ds\zeta}{R_\sigma} \frac{1 - (R_c / R_\sigma)^{1/2}}{1 + (R_c / R_\sigma)^{1/2}} \approx \frac{Ds\zeta}{R_\sigma}. \tag{2.11}$$

Thus, in the case of the initial condition in the form of (2.3), the bubble growth rate in the end of the first stage increases twice in comparison with the initial value.

The second stage of the process will be when the bubble growth occurs within the interval of sizes

$$(R_\sigma R_c)^{1/2} \leq R \leq R_\sigma. \tag{2.12}$$

During all this stage, as $(R_c R_\sigma)^{1/2} \gg R_c$, it is already valid that $R \gg R_c$, thus the driving force $n_0 - n_R$ remains practically constant. As a result, bubble growth decelerates, although, as it was noted above, the counteraction of Laplace forces to the growth gradually is attenuated. The bubble growth rate, following (2.1), decreases during the second stage from its maximal value (2.11) in the beginning of the stage to the value

$$\dot{R}\Big|_{R=R_\sigma} = \frac{Ds\zeta}{2R_\sigma}(1 - R_c/R_\sigma) \approx \frac{Ds\zeta}{2R_\sigma}, \qquad (2.13)$$

decreasing two-fold and coming back to the initial value (2.10).

The Laplace forces contribution $2\sigma/R$ to the pressure inside the bubble increases during the second stage by $(R_\sigma/R_c)^{1/2}$ times and at the completion of this stage becomes comparable with the external pressure $\Pi$ contribution (from (1.13) we have strict equality $\frac{2\sigma}{R_\sigma} = \frac{3}{2}\Pi$). It is this physical condition that defines the completion of the second stage.

On the subsequent, third stage, which corresponds to the interval of sizes $R \geq R_\sigma$, monotonous decelerated bubble growth continues. At the same time, the role of Laplace forces continues to decrease gradually, and the pressure inside the bubble approaches to a constant value equal to the external pressure $\Pi$. As it will be shown in the next section, this process is rather protracted, so the concluding phase of the third stage, when the pressure inside the bubble practically does not change and the use of self-similar solution [16] is possible, comes only in the interval of sufficiently large sizes of the bubble, when the condition $R \gg R_\sigma$ is satisfied with a certain reserve.

Concluding this section we will study the condition of applicability of steady approximation (1.10) expressed as a strong inequality (1.17) on each of the consecutive stages of bubble growth. As it was noted in section 1, limitation (1.17) becomes stricter with the increase of $R$. In the interval of extremely large bubble sizes (the concluding phase of the third stage) this limitation transforms into the strong inequality

$$s^{1/2} \ll 1/\zeta^{1/2}. \qquad (2.14)$$

Obviously, when condition (2.14) is fulfilled, the applicability of steady approximation is valid during the whole process of bubble growth. If initial solution parameters are such that condition (2.14) is not fulfilled, then the steady condition will be violated as the bubble size increases. Thus transition to non-steady growth regime is inevitable. The size at which such a transition takes place depends on the values of parameters $s$ and $\zeta$.

Let us clarify the limitations on $s$ and $\zeta$ imposed by condition (1.17) during each of the consecutive stages of bubble growth.

Steady state of bubble growth during the first stage takes place, if inequality (1.17) is fulfilled at $R = (R_c R_\sigma)^{1/2}$. Corresponding limitation is expressed (let us remind, that $R_\sigma \gg R_c$) by inequality $(s\zeta)^{1/2} \ll (R_\sigma / R_c)^{1/4}$, which, accounting for equality (1.14), can be presented as:

$$s^{1/2} \ll \left(\frac{2}{3\zeta}\right)^{1/4}. \qquad (2.15)$$

It is evident that inequality (2.15) is significantly weaker than limitation (2.14). In order to guarantee the condition of steady bubble growth during the second stage, it is necessary to fulfill it at $R = R_\sigma$. As it follows from inequality (1.17), with account for the relation $R_\sigma \gg R_c$ this requirement leads to the limitation

$$s^{1/2} \ll \left(\frac{2}{\zeta}\right)^{1/2}. \qquad (2.16)$$

Steady growth on the third stage, when the bubble size can be arbitrarily large, takes place when the condition (2.14) is fulfilled. Let us note that limitations (2.14) and (2.16) practically coincide.

In accordance with inequality (1.15), supersaturation $\zeta$ should be significant. Using this fact in obtained limitations (2.14) – (2.16) we have that steady approximation for the bubble growth is applicable at all the stages only in the case of extremely small gas solubility. The weakest limitation is the one needed during the first stage of growth. Nevertheless, at supersaturation values under consideration $\zeta \gg 10$, from (2.15) it follows that even during this stage the steady approximation could be used only in the case when gas solubility is rather small.

Let us note that in the very beginning of the process of the far supercritical bubble growth, i. e. in the size interval $0 \leq R - 2R_c \ll R_c$, the steady condition (1.17) can be expressed as the following inequality with account for inequality $R_\sigma \gg R_c$ and equality

(1.14)

$$s^{1/2} \ll \left(\frac{2}{3}\right)^{1/2}, \qquad (2.17)$$

which does not include large quantity $\zeta$. Limitation (2.17) is significantly weaker than inequalities (2.14) – (2.16), and thus it is well justified, in accordance with the tabular information from [19, 20] on solubility of different gases in different liquids. Following this tabular data it follows that $10^{-2} < s < 10^{-1}$ (in the absence of dissociation and chemical transitions of dissolved molecules).

## 3. DURATION OF CONSECUTIVE STAGES AND TREND TOWARDS SELF-SIMILAR REGIME OF BUBBLE GROWTH

Let us write the bubble growth equation (1.12), i. e. equation (2.1), in the form which is convenient for integration:

$$R\dot{R} + (R_\sigma + R_c)\dot{R} + (R_\sigma + R_c)R_c \frac{\dot{R}/R_c}{R/R_c - 1} = Ds\zeta. \qquad (3.1)$$

Integrating equation (3.1), we obtain

$$\frac{R^2}{2} + (R_\sigma + R_c)R + (R_\sigma + R_c)R_c \ln\left(\frac{R}{R_c} - 1\right) = Ds\zeta(t + \tau), \qquad (3.2)$$

where $\tau$ is the constant which has time dimensionality and is defined by the initial value of radius at time $t = 0$. Using initial condition (2.3), from equation (3.2) we find

$$\tau = \frac{2R_c(R_\sigma + 2R_c)}{Ds\zeta}. \qquad (3.3)$$

Excluding time $\tau$ from equation (3.2) by means of equation (3.3), we obtain

$$\frac{R^2 - 4R_c^2}{2} + (R_\sigma + R_c)(R - 2R_c) + (R_\sigma + R_c)R_c \ln\left(\frac{R}{R_c} - 1\right) = Ds\zeta t \quad (R \geq 2R_c). \qquad (3.4)$$

In equality (3.4) instead of limitation $R \geq 2R_c$, accounting for initial condition (2.3), we could write limitation $t \geq 0$.

Validity of the general equation (3.4), which accounts strictly and analytically for Laplace forces influence on the bubble growth process, is limited only by the condition of applicability of steady approximation (1.10) for the diffusion flux. Equation (3.4) complies with the results obtained in papers [5, 6] for the particular case of steady growth of a bubble. In papers [5, 6] it was not clarified what the limitation on the bubble radius is when these formulas obtained without account for the fluctuation influence on bubble growth, are valid.

Equation (3.4) does not imply the smallness of quantity $R_c/R_\sigma$ which follows from relations (1.14) and (1.15). As it was shown in the previous section, when this ratio is small, one can, using bubble growth equation (2.1) and initial condition (2.3), trace three characteristic stages in the bubble evolution after its nucleation. Equation (3.4), giving the explicit dependence of bubble radius on time, allows us, in particular, to find characteristic times corresponding to consecutive stages.

According to relations (2.4) and (2.8), at the end of the first stage the bubble radius reaches, the value $R_m = (R_c R_\sigma)^{1/2}$. Substituting value $R = R_m$ to the equation (3.4) and considering that, (by virtue of inequalities $R_c \ll R_m \ll R_\sigma$), the main contribution to the left hand side of (3.4) is made by the second addend, we obtain the expression for the first stage duration $t_1$

$$t_1 = \frac{R_\sigma^2}{Ds\zeta}\left(\frac{R_c}{R_\sigma}\right)^{1/2}. \tag{3.5}$$

Using equation (1.14), expression (3.5) can be also presented in the form

$$t_1 = \left(\frac{3}{2}\right)^{1/2}\frac{R_\sigma^2}{Ds\zeta^{3/2}}. \tag{3.6}$$

As it follows from (3.6), with the increase of initial supersaturation of the solution, the first stage duration decreases proportionally to $1/\zeta^{3/2}$.

The second stage of bubble growth starts at the time point $t_1$ and finishes at the time point $t_2$ defined by the condition $R|_{t=t_2} = R_\sigma$. Substituting value $R = R_\sigma$ to the equation (3.4) and considering that, by virtue of inequality $R_c \ll R_\sigma$, the main contribution to the left hand side of (3.4) is made by the first and the second addends, we obtain the ex-

pression for $t_2$

$$t_2 = \frac{3}{2}\frac{R_\sigma^2}{Ds\zeta}. \tag{3.7}$$

As one can see from this expression, $t_2$ dependence on initial solution supersaturation is defined by multiplier $1/\zeta$. The duration of the second stage is much longer than the duration of the first stage, as from expressions (3.6) and (3.7) it follows that

$$\frac{t_2}{t_1} = \left(\frac{3\zeta}{2}\right)^{1/2} \gg 1. \tag{3.8}$$

Let us note, that the time dependence of the bubble radius defined by relation (3.4) has an especially simple character (direct proportionality) in the interval $R_c \ll R \ll R_\sigma$. Indeed, in this interval the main contribution to the left hand side of the equation (3.4) is made only by the second addend. Thus from (3.4) we obtain

$$R = \frac{Ds\zeta}{R_\sigma}t \qquad \left(\frac{R_\sigma R_c}{Ds\zeta} \ll t \ll \frac{R_\sigma^2}{Ds\zeta}\right). \tag{3.9}$$

Before we approach the consideration of the third stage of bubble growth, let us return to the question about the choice of initial condition in the form (2.3). Let us begin the description of the bubble growth using equation (2.1) from the size $R = R_c + (kT/4\pi\sigma)^{1/2} \approx 1.1R_c$, instead of the interval $R \geq 2R_c$, where, according to section 2, the smallness of fluctuations contributions is automatically guaranteed. Let us find the time $\Delta t$ of bubble growth from the size $R = 1.1R_c$ to the size $R = 2R_c$ and show that, at the values of system parameters under consideration, this time is small even in comparison with time $t_1$ – duration of the first and shortest stage of growth.

Substituting in equation (3.2) $R = 1.1R_c$ and $R = 2R_c$ and subtracting the obtained equations, we have:

$$1.4R_c^2 + 3,2R_\sigma R_c = Ds\zeta\Delta t. \tag{3.10}$$

Accounting for the strong inequality $R_\sigma \gg R_c$, we obtain the following from equation (3.10)

$$\Delta t = 3.2 \frac{R_\sigma R_c}{Ds\zeta} . \qquad (3.11)$$

From (3.5), (3.11) and (1.14) the following evident equation for the ration of the time $\Delta t$ and $t_1$ can be concluded:

$$\frac{\Delta t}{t_1} = 3.2 \left(\frac{3}{2\zeta}\right)^{1/2} . \qquad (3.12)$$

At the characteristic supersaturation value $\zeta \sim 10^3$ formula (3.12) gives us $\Delta t/t_1 \sim 0.1$.

Now let us consider the third stage of bubble growth, when $R \geq R_\sigma$ and $t \geq t_2$. First of all, let us note that at the end of the first stage, when the bubble radius $R$ approaches the value $(R_c R_\sigma)^{1/2}$, and with even more assurance on the second and the third stages, one can neglect the logarithmic addend in equation (3.4) (it was accounted for earlier, when expressions (3.5), (3.7) and (3.9) were obtained). Indeed, at $R \geq R_\sigma$ the ratio between the third addend in (3.4) and the second one is a small value in accordance with inequality

$$\frac{R_c}{R} \ln\left(\frac{R}{R_c}\right) \ll 1 . \qquad (3.13)$$

Moreover, since during the third stage the main contribution in the left hand side of (3.4) gradually with increase of the bubble radius $R$ tends to come from the first addend equal to $R^2/2$; and the contribution of the second addend (influence of Laplace forces) decreases, neglect of logarithmic contribution becomes fairly justified. As a result, equation (3.4) conformably to the third stage of bubble growth can be written in the form of

$$\frac{R^2}{2} + R_\sigma R = Ds\zeta t . \qquad (3.14)$$

Solving quadratic equation (3.14) on $R$, we find explicit dependence of bubble radius on time during the third stage of growth:

$$R = (2Ds\zeta t + R_\sigma^2)^{1/2} - R_\sigma \qquad (t \geq t_2). \qquad (3.15)$$

The obtained expression is valid at all $t \geq t_2$ only when condition (2.14), guaranteeing the assumed steadiness of the bubble growth during this stage, is fulfilled.

Assuming this condition to be fulfilled, we use relation (3.15) for the description of the trend of the third stage towards self-similar regime of bubble growth studied in [16]. According to [16], self-similar regime of growth requires that the bubble has such a large size that the influence of Laplace forces can be excluded from consideration. However, this regime of growth does not imply any limitations on gas solubility. As it has been shown in [16], while we consider conditions of small gas solubility, self-similar regime is steady when limitation $(s\zeta)^{1/2} \ll 1$ is fulfilled, which is in exact concordance with condition (2.14).

The time dependence of the bubble radius obtained using self-similar solution [16] in the limiting case $(s\zeta)^{1/2} \ll 1$ is given by expression

$$R_* = (2Ds\zeta t)^{1/2} \quad (t \geq t_0), \qquad (3.16)$$

where time $t$ is counted, as in the present paper, from the moment of nucleation of the far supercritical bubble. Symbol * at $R$ indicates that the solution was obtained within the framework of the self-similar theory. Characteristic time $t_0$ in the dependence (3.16) is chosen in such a way that the bubble radius at this time significantly exceeds the value $2\sigma/\Pi$:

$$R_0 \gg \frac{2\sigma}{\Pi}, \qquad (3.17)$$

where radius $R_0$ is defined by the condition

$$R_0 = R_*(t)\big|_{t=t_0}. \qquad (3.18)$$

Then, from (3.16) it follows that

$$t_0 = \frac{R_0^2}{2Ds\zeta}. \qquad (3.19)$$

In accordance with inequality (3.17), in the bubble with radius $R_0$ the contribution of Laplace forces to the pressure is significantly less than the value of external pressure.

Let us note that within the framework of self-similar theory it is impossible to make a strict evaluation of the accuracy of time dependence of bubble radius obtained using this theory. In [16] it has been assumed that at any arbitrary value of solubility the relative accuracy of expression (3.16) is equal to the relative deviation of the pressure in the bubble from the external pressure; and applicability of expression (3.16) has been determined using the estimation

$$R_0 \cong \frac{40\sigma}{\Pi} \tag{3.20}$$

(estimation (3.8) in [16]) which guarantees the justification of inequality (3.17) with a considerable reserve.

The obtained expression (3.15) which accounts for the influence of Laplace forces strictly, allows us to make a strict evaluation of the accuracy of self-similar approximation in the considered case $(s\zeta)^{1/2} \ll 1$, when the steady growth of a gas bubble takes place.

From equations (3.20) and (1.13) it follows that

$$R_0/R_\sigma \cong 30. \tag{3.21}$$

Correspondingly from equations (3.19) and (3.7) we find

$$t_0/t_2 = \frac{1}{3}\left(\frac{R_0}{R_\sigma}\right)^2 \cong 3\times 10^2. \tag{3.22}$$

As it follows from equation (3.22), strict equation (3.15) is applicable *a fortiori* at time $t \geq t_0$. At the same time, using equality (3.16), we can write equation (3.15) in the following form

$$R = (R_*^2 + R_\sigma^2)^{1/2} - R_\sigma \quad (t \geq t_0). \tag{3.23}$$

By virtue of relation (3.21) and inequality $R_* \geq R_0 \quad (t \geq t_0)$, we have

$$R_\sigma/R_* < 1/30 \quad (t \geq t_0), \tag{3.24}$$

thus $R_\sigma/R_* \ll 1 \quad (t \geq t_0)$. Now from equation (3.23) now we obtain

$$R(t) = R_*(t) - R_\sigma[1 - R_\sigma/2R_*(t)] \quad (t \geq t_0), \tag{3.25}$$

where we explicitly indicate the dependence of $R$ and $R_*$ on $t$.

Equation (3.25) gives us the analytical description of the trend of the exact solu-tion (3.15) towards the self-similar regime of bubble growth at times $t \geq t_0$. As it follows from the expression (3.25) which, as can be seen from (3.24), is practically equivalent to the expression

$$R_*(t) = R(t) + R_\sigma[1 - R_\sigma/2R(t)] \qquad (t \geq t_0), \tag{3.26}$$

self-similar solution $R_*(t)$ gives us overestimated value of bubble radius. The difference between $R_*(t)$ and exact solution $R(t)$ increases monotonically with time, approaching constant value $R_\sigma$, and for extremely large $t$ we have $R_*(t) - R(t) \approx R_\sigma$. For the relative deviation of $R_*(t)$ from $R(t)$ caused by the influence of Laplace forces from equation (3.25) we find:

$$\frac{R_*(t) - R(t)}{R_*(t)} = \frac{R_\sigma}{R_*(t)}[1 - R_\sigma/2R_*(t)] \qquad (t \geq t_0). \tag{3.27}$$

In accordance with inequality (3.24), this deviation does not exceed $1/30$ at $t = t_0$. Thus, strictly defined deviation (3.27) turns out to be less than the evaluation obtained in [16], which was calculated without the use of condition of small gas solubility. With the further increase of $t$ the relative deviation decreases, as it can be easily proven using equation (3.16) proportional to $(t/t_0)^{-1/2}$.

Trend towards the concluding phase of the third stage of bubble growth is, by virtue of equations (3.22) and (3.8), an extremely protracted process in comparison with the bubble growth during the first two stages. Thus, the "memory" of the influence of Laplace forces in the time dependence of the bubble radius remains for a relatively long time.

## CONCLUSIONS

The discussion presented above has shown that at high values of initial supersaturation of solution necessary for the formation of supercritical nucleus and real values of solubility the gas bubble growth is a multistage process and, as a rule, it turns into nonsteady regime. The influence of Laplace forces remains valid for a long period of bub-ble growth.

Thus, it is crucial to obtain the solution of the problem of physically correct description of non-steady growth of a gas bubble in the interval of its sizes $R \geq 2R_c$ with account for the influence of Laplace forces and high value of initial supersaturation. The conditions and regularities of the steady bubble growth on each of the revealed stages, including at $0 \leq R - 2R_c \ll R_c$, give reliable foundations to obtain a solution of such problem.

This work was supported by the Analytical Program "The Development of Scientific Potential of Higher Education" (2009-2010), project RNP.2.1.1.4430 Structure, Thermodynamics and Kinetics of Supramolecular Systems. Research by G. Yu. Gor was also supported by Postgraduate scholarship by K. I. Zamaraev fund.

12. *Slezov V. V., Abyzov A. S., Slezova Zh. V.* // Colloid Journal 2004. V. 66. P. 643.
13. *Slezov V. V., Abyzov A. S., Slezova Zh. V.* // Colloid Journal 2005. V. 67. P. 94.
14. *Chernov A. A. //* J. Appl. Mech. Tech. Phys. 2003 V. 44 P. 80.
15. *Chernov A. A., Kedrinskii, V. K., Davidov M. N.* // J. Appl. Mech. Tech. Phys. 2004. V. 45. P. 162.
16. *Grinin A. P., Kuni F. M., Gor G. Yu.* // Colloid Journal. 2009. V. 71. P. 47. http://arxiv.org/abs/0808.3706v1
17. *Landau L. D., Lifshits E. M.* Statistical Physics. Part 1, Oxford: Pergamon Press, 1980.
18. *Lifshits E. M., Pitaevsky L. P.* Physical Kinetics, Oxford: Pergamon Press, 1981.
19. Chemist's Handbook (*Spravochnik Khimika* in Russian) V. 3 / ed. by Nikolsky B. P. et al., Moscow: GNTI Khimicheskoy literatury, 1963.
20. Brief handbook of physico-chemical values (*Kratkiy physico-khimicheskiy spravochnik* in Russian) / ed by Mishchenko K. P. and Ravdel A. A., Leningrad: Khimia, 1974.
22